\documentclass[10pt,conference]{IEEEtran}
\usepackage{amsmath}
\usepackage{graphicx}
\begin{document}

\title{Hierarchical Decoupling Principle of a MIMO-CDMA Channel 
in Asymptotic Limits}

\author{
\authorblockN{Keigo Takeuchi}
\authorblockA{Graduate School of Informatics \\
Kyoto University \\
Kyoto, Japan \\
takeuchi@sys.i.kyoto-u.ac.jp}
\and
\authorblockN{Toshiyuki Tanaka}
\authorblockA{Graduate School of Informatics \\
Kyoto University \\
Kyoto, Japan \\
tt@i.kyoto-u.ac.jp}
}
% avoiding spaces at the end of the author lines is not a problem with
% conference papers because we don't use \thanks or \IEEEmembership
%Homer Simpson\authorrefmark{2},
%Montgomery Scott\authorrefmark{3} and
%Eldon Tyrell\authorrefmark{4}}
%Telephone: (800) 555--1212, Fax: (888) 555--1212}
%\authorblockA{\authorrefmark{4}Tyrell Inc., 123 Replicant Street, Los Angeles, California 90210--4321}}

\maketitle

\begin{abstract}
We analyze an uplink of a fast flat fading MIMO-CDMA channel in the case 
where the data symbol vector for each user follows an arbitrary 
distribution. The spectral efficiency of the channel with CSI at the 
receiver is evaluated analytically with the replica method. 
The main result is that the hierarchical decoupling principle holds 
in the MIMO-CDMA channel, i.e., the MIMO-CDMA channel is decoupled 
into a bank of single-user MIMO channels in the many-user limit, 
and each single-user MIMO channel is further decoupled into a bank of 
scalar Gaussian channels in the many-antenna limit for a fading 
model with a limited number of scatterers.  
\end{abstract}

\section{Introduction}
%CDMA
Direct-sequence spread-spectrum code-division multiple access (CDMA) 
has been utilized as a multiple access scheme in wireless communication. 
%MIMO
As a method of overcoming the capacity bottleneck in future  
wireless communication, multiple-input 
multiple-output (MIMO) systems have attracted so much attention 
since the latter half of the 1990s \cite{Tse05}. 
Recently, MIMO systems with CDMA technology (MIMO-CDMA) have been 
studied \cite{Mantravadi03, Ni05}. 

%purpose
Mantravadi et al. \cite{Mantravadi03} evaluated the asymptotic 
spectral efficiency of a MIMO-CDMA channel with Gaussian 
modulation using the random matrix theory. However, other data 
modulation schemes such as quadrature phase shift keying (QPSK) 
modulation, or more generally, $M$-quadrature amplitude modulation 
($M$-QAM) are commonly employed in practice. 
Thus, it is important to analyze the MIMO-CDMA channel with a 
modulation of the kind. The purpose of our study is to evaluate 
the spectral efficiency of the MIMO-CDMA channel in the case that 
the data symbol vector for each user follows an arbitrary distribution.     

%replica method
It was reported \cite{Tanaka02} that the spectral efficiency of a 
CDMA channel with binary phase shift keying modulation can be evaluated 
with the so-called replica method. Then, the method was applied to the  
performance evaluation of MIMO systems \cite{Moustakas03,Muller03,Wen06}.   
%decoupling 
Guo et al. \cite{Guo05} analyzed a CDMA channel in the case 
that the data symbol for each user follows an arbitrary distribution 
and claimed that the decoupling principle, the equivalent result to which 
has already been proved in the case of Gaussian modulation 
\cite{Tse99, Verdu99}, holds in an asymptotic limit, i.e., 
the CDMA channel is decoupled into a bank of scalar Gaussian channels. 
This principle makes possible the analytical evaluation of the 
spectral efficiency since the degree of freedom drastically decreases.   
However, it is not still clear except special cases \cite{Mantravadi03} 
whether or not the decoupling principle holds in the MIMO-CDMA channel.  

In this paper, we claim that the decoupling principle holds in the 
MIMO-CDMA channel and evaluate the spectral efficiency of the channel 
using the replica method. 

\section{Model}
We consider the uplink of a synchronous $K$-user MIMO-CDMA flat 
fading channel \cite{Mantravadi03}  
\begin{equation} \label{MIMO_CDMA_channel}
\boldsymbol{y}_{l} = 
\sum_{k=1}^{K}s_{l}^{k}\boldsymbol{H}^{k}\boldsymbol{x}^{k} 
+ \boldsymbol{n}_{l}, 
\end{equation}
where the $k$th user has $M_{k}$ transmit antennas and the receiver 
has $N$ receive antennas. 
$\boldsymbol{x}^{k}=(x_{1}^{k}, \ldots, x_{M_{k}}^{k})^{T}$ is 
the data symbol vector for the $k$th user and 
$\{s_{l}^{k};l=1,\ldots, L\}$ is the spreading sequence 
for the $k$th user. 
We assume that $\{\boldsymbol{x}^{k};k=1,\ldots, K\}$ 
are mutually independent complex random variables and that 
the real part and the imaginary part of $\{s_{l}^{k}\}$ are 
independent and identically distributed (i.i.d.) 
zero-mean random variables with the variance $1/2L$.  
$\boldsymbol{H}^{k}$ represents the $N$-by-$M_{k}$ channel matrix 
for the $k$th user, i.e., 
the ($n$, $m_{k}$)-element $h_{nm_{k}}^{k}$ of $\boldsymbol{H}^{k}$ 
is the channel gain  from the $m_{k}$th transmit antenna of the 
$k$th user to the $n$th receive antenna. 
We assume that $\{\boldsymbol{H}^{k};k=1,\ldots, K\}$ 
are mutually independent. 
We consider the case that the noise is additive white Gaussian noise 
(AWGN), i.e., $\{\boldsymbol{n}_{l}; l=1,\ldots, L\}$ are i.i.d. 
zero-mean circularly symmetric complex Gaussian random variables 
with the covariance matrix $\sigma^{2}\boldsymbol{I}_{N}$, 
which are denoted by $\mathcal{CN}(\boldsymbol{0}, 
\sigma^{2}\boldsymbol{I}_{N})$.   
$\boldsymbol{y}_{l}$ represents the $N$-dimensional received signal vector. 

We write the entire data symbol vector $\vec{\boldsymbol{x}}$ and 
the received signal vector $\vec{\boldsymbol{y}}$ in a symbol period as 
$\vec{\boldsymbol{x}}=({\boldsymbol{x}^{1}}^{T}, \ldots , 
{\boldsymbol{x}^{K}}^{T})^{T}$, and $\vec{\boldsymbol{y}}=
({\boldsymbol{y}_{1}}^{T}, \ldots , {\boldsymbol{y}_{L}}^{T})^{T}$, 
respectively. 
The maximal sum rate in the fast fading 
channel with perfect channel side information at the receiver 
is given by the conditional mutual 
information (the base of logarithms is taken to $2$ in this paper) 
between $\vec{\boldsymbol{x}}$ and $\vec{\boldsymbol{y}}$ 
conditioned on the spreading sequences and the channel matrices 
\cite{Tse05}   
\begin{equation}  \label{mutual_information}
I(\vec{\boldsymbol{x}};\vec{\boldsymbol{y}}| \boldsymbol{\mathcal S}, 
\boldsymbol{\mathcal H}) = 
\mathrm{E}\left[
  \log\frac{
   p(\vec{\boldsymbol{y}}|\vec{\boldsymbol{x}}, 
   \boldsymbol{\mathcal S}, \boldsymbol{\mathcal H})
  }
  {
   \mathrm{E}_{\vec{\boldsymbol{x}}}\left[
    p(\vec{\boldsymbol{y}}| \vec{\boldsymbol{x}}, 
    \boldsymbol{\mathcal S},\boldsymbol{\mathcal H})
   \right]
  }  
\right],
\end{equation}
where ${\boldsymbol{\mathcal S}}=
\{s_{l}^{k}; l=1,\ldots, L, k=1, \ldots , K\}$, 
${\boldsymbol{\mathcal H}}=\{\boldsymbol{H}^{k}; k=1, \ldots , K\}$, 
and where $p(\vec{\boldsymbol{y}}|\vec{\boldsymbol{x}}, 
\boldsymbol{\mathcal S}, \boldsymbol{\mathcal H})$ is given by
\begin{equation} 
p(\vec{\boldsymbol{y}}|\vec{\boldsymbol{x}}, \boldsymbol{\mathcal S}, 
\boldsymbol{\mathcal H}) = 
\frac{1}{(\pi\sigma^{2})^{LN}}\prod_{l=1}^{L}
\mathrm{e}^{
 -\frac{1}{\sigma^{2}}
 \left\|
  \boldsymbol{y}_{l} 
  - \sum_{k=1}^{K}s_{l}^{k}\boldsymbol{H}^{k}\boldsymbol{x}^{k}
 \right\|^{2}
}. 
\end{equation} 
We define the spectral efficiency $\mathcal{C}_{\mathrm{MIMO-CDMA}}$ 
as the maximal sum rate per chip and per transmit antenna.  
In the many-user limit, where the number of users $K$ and the spreading 
factor $L$ tend to infinity with their ratio $\beta=K/L$ fixed, 
the spectral efficiency is given by 
\begin{equation} \label{spectral_efficiency}
{\mathcal C}_{\mathrm{MIMO-CDMA}} = 
\frac{\beta}{\bar{M}\ln 2}\mathrm{E}_{\boldsymbol{\mathcal H}}\left[
 {\mathcal F}
\right]
- \bar{\mu}^{-1}\log(\pi\sigma^{2}\mathrm{e}), 
\end{equation}  
where $\bar{M}=\lim_{K\rightarrow\infty}K^{-1}\sum_{k=1}^{K}M_{k}$ and  
$\bar{\mu}$ is the ratio of the average number of transmit antennas to 
the number of receive antennas, i.e., $\bar{\mu}=\bar{M}/N$. 
The {\it free energy} $\mathcal F$ is defined as
\begin{equation}
{\mathcal F} = 
-\lim_{K,L\rightarrow \infty}\frac{1}{K}
\mathrm{E}
\left[
 \left.
 \ln p(\vec{\boldsymbol{y}}| \boldsymbol{\mathcal S}, 
 \boldsymbol{\mathcal H})
 \right| \boldsymbol{\mathcal{H}} 
\right]. \label{free_energy} 
\end{equation}  

\section{Replica Analysis of MIMO-CDMA}
We explain briefly the calculation procedure of the free 
energy~(\ref{free_energy}) \cite{Tanaka02,Guo05}. 
Substituting the following identity:
\begin{equation}
\lim_{u\rightarrow 0}\frac{\partial}{\partial u}
\left[
 p(\vec{\boldsymbol{y}}| \boldsymbol{\mathcal S},\boldsymbol{\mathcal H})
\right]^{u} 
= 
\ln p(\vec{\boldsymbol{y}}| \boldsymbol{\mathcal S}, 
\boldsymbol{\mathcal H})
\end{equation}
to (\ref{free_energy}), we obtain 
\begin{equation} \label{free_energy_tmp}
{\mathcal F} = 
-\lim_{K,L\rightarrow \infty}
\lim_{u\rightarrow 0}\frac{\partial}{\partial u}\Xi_{K}^{(u)},
\end{equation}
\begin{equation}
\Xi_{K}^{(u)} = \frac{1}{K}\ln\mathrm{E}\left\{ 
 \left[
  p(\vec{\boldsymbol{y}}| \boldsymbol{\mathcal S}, 
  \boldsymbol{\mathcal H})
 \right]^{u} 
 | \boldsymbol{\mathcal{H}}
\right\}.
\end{equation}
We assume that the limit with respect to $K$, $L$ and the operation 
with respect to $u$ are interchangeable. 
Then, (\ref{free_energy_tmp}) becomes  
\begin{equation}
{\mathcal F} = 
-\lim_{u\rightarrow 0}\frac{\partial}{\partial u}
\lim_{K,L\rightarrow \infty}\Xi_{K}^{(u)}. \label{free_energy_replica}
\end{equation}
Further, we assume that the result for positive integers $u$ 
is valid for real number $u$ in the calculation of 
(\ref{free_energy_replica}). 

We define $N$-dimensional vectors $\boldsymbol{v}^{\alpha}$ as 
\begin{equation}
\boldsymbol{v}^{\alpha} = 
\frac{1}{\sqrt{\beta}}\sum_{k=1}^{K}s^{k}\boldsymbol{H}^{k}
\boldsymbol{x}^{k,\alpha}, 
\quad 
\alpha = 0, \ldots , u, 
\end{equation}
where $\boldsymbol{x}^{k,0}$ and $\boldsymbol{x}^{k,\alpha}=
(x_{1}^{k,\alpha}, \ldots, x_{M_{k}}^{k,\alpha})^{T}$ are 
the original data symbol vector and the replicated data symbol vector 
for the $k$th user, respectively.  
$\{\boldsymbol{x}^{k,\alpha},\alpha=0,\ldots,u\}$ are i.i.d. random 
variables following $p(\boldsymbol{x}^{k})$ and  
$\{s^{k};k=1, \ldots, K\}$ are i.i.d. random variables following 
$p(s_{l}^{k})$.
Then, the expectation of $[p(\vec{\boldsymbol{y}}| \boldsymbol{\mathcal S}, 
\boldsymbol{\mathcal H})]^{u}$ with respect to $\vec{\boldsymbol{y}}$  
and $\boldsymbol{\mathcal S}$ is given by 
\begin{equation} \label{tmp1}
\mathrm{E}
\left\{
 [p(\vec{\boldsymbol{y}}| \boldsymbol{\mathcal S}, 
 \boldsymbol{\mathcal H})]^{u}
 | \boldsymbol{\mathcal{H}} 
\right\} = 
\mathrm{E}\left\{
 \left.
 \exp\left[
  LG_{K}^{(u)}(\vec{\boldsymbol{\mathcal X}},\boldsymbol{\mathcal{H}})
 \right]
 \right| \boldsymbol{\mathcal{H}}  
\right\}, 
\end{equation}
\begin{eqnarray} \label{G_K}
G_{K}^{(u)}(\vec{\boldsymbol{\mathcal X}}, \boldsymbol{\mathcal{H}}) =  
\ln \mathrm{E}_{\boldsymbol{S}}\left[
  \int\prod_{\alpha =0}^{u}\mathrm{e}^{
   -\frac{1}{\sigma^{2}}\left\|
    \boldsymbol{y} 
    - \sqrt{\beta}\boldsymbol{v}^{\alpha} 
   \right\|^{2}
  } 
  \boldsymbol{dy} 
\right] \nonumber \\
-N(u+1)\ln(\pi\sigma^{2}),
\end{eqnarray}
where $\vec{\boldsymbol{\mathcal X}}$ is defined as 
$\{\boldsymbol{x}^{k,\alpha}; k=0,\ldots,K, \alpha =0, \ldots ,u\}$
and $\boldsymbol{S}$ repserents $\{s^{k}; k=1, \ldots, K\}$. 
When $K$ and $L$ are sufficiently large with their ratio fixed, 
due to the central limit theorem,  
$\boldsymbol{v} = ({\boldsymbol{v}^{0}}^{T}, \ldots , 
{\boldsymbol{v}^{u}}^{T})^{T}$ conditioned on 
$\vec{\boldsymbol{\mathcal X}}$ and $\boldsymbol{\mathcal H}$ 
follows approximately the zero-mean circularly symmetric complex Gaussian 
distribution with the covariance matrix 
\begin{equation} 
\boldsymbol{\mathcal Q} = 
\frac{1}{K}\sum_{k=1}^{K}\boldsymbol{w}^{k}{\boldsymbol{w}^{k}}^{*}, 
\end{equation}
where $\boldsymbol{w}^{k}$ is defined as 
$[(\boldsymbol{H}^{k}\boldsymbol{x}^{k,0})^{T}, \ldots,   
(\boldsymbol{H}^{k}\boldsymbol{x}^{k,u})^{T}]^{T}$. Then, we can evaluate   
(\ref{tmp1}) and (\ref{G_K}) as    
\begin{equation}
\Xi_{K}^{(u)} = 
\frac{1}{K}\ln\mathrm{E}\left\{
 \mathrm{e}^{ 
  K\beta^{-1}G^{(u)}(\boldsymbol{\mathcal Q})
 }
\right\} 
+ \mathcal{O}(K^{-1}), \label{tmp2} 
\end{equation}
\begin{equation}
G^{(u)}(\boldsymbol{\mathcal Q}) = 
-\ln\det(\boldsymbol{I}+\boldsymbol{\Sigma}\boldsymbol{\mathcal Q}) 
-Nu\ln (\pi\sigma^{2}) 
-N\ln(1+u), \label{G}
\end{equation}
where $\boldsymbol{\Sigma}$ is defined as 
\begin{equation}
\boldsymbol{\Sigma} = 
\frac{\beta}{\sigma^{2}(1+u)}
\begin{bmatrix}
 u & -\boldsymbol{e}_{u}^{T} \\
 - \boldsymbol{e}_{u} & (1+u)\boldsymbol{I}_{u} 
 - \boldsymbol{e}_{u}\boldsymbol{e}_{u}^{T}  
\end{bmatrix}
\otimes \boldsymbol{I}_{N}, 
\end{equation}
where $\boldsymbol{e}_{u}$ is the $u$-dimensional vector whose 
elements are all one and $\otimes$ represents the Kronecker product.

With a Hermitian matrix $\tilde{\boldsymbol{\mathcal Q}}$ we define 
the moment generating function of the data symbols of the $k$th user as  
\begin{equation}
{\mathcal M}_{k}^{(u)}(\tilde{\boldsymbol{\mathcal Q}}) =
\mathrm{E}_{\boldsymbol{\mathcal X}^{k}}\left\{
 \exp\left[
  \mathrm{tr}\left(
   \tilde{\boldsymbol{\mathcal Q}}\boldsymbol{w}^{k}{\boldsymbol{w}^{k}}^{*} 
  \right)
 \right]
\right\},  \label{moment_generating}
\end{equation} 
where $\boldsymbol{\mathcal X}^{k}$ represents  
$\{x_{m_{k}}^{k,\alpha}; m_{k}=1, \ldots, M_{k}, \alpha=0, \ldots, u\}$. 
Since $\boldsymbol{\mathcal{Q}}$ satisfies the large deviation principle,  
with the saddle point method, (\ref{tmp2}) is evaluated as
\begin{equation} \label{tmp3}
\lim_{K,L\rightarrow\infty}\Xi_{K}^{(u)} = 
\sup_{\boldsymbol{\mathcal Q}}\left[
 \beta^{-1}G^{(u)}(\boldsymbol{\mathcal Q}) 
 - I^{(u)}(\boldsymbol{\mathcal Q})
\right], 
\end{equation} 
where the rate function $I^{(u)}(\boldsymbol{\mathcal Q})$ is given by 
\begin{equation}
I^{(u)}(\boldsymbol{\mathcal Q}) = 
\sup_{\tilde{\boldsymbol{\mathcal Q}}}\left[
 \mathrm{tr}(\tilde{\boldsymbol{\mathcal Q}}\boldsymbol{\mathcal Q}) 
 - \lim_{K\rightarrow\infty}\frac{1}{K}\sum_{k=1}^{K}
 \ln{\mathcal M}_{k}^{(u)}(\tilde{\boldsymbol{\mathcal Q}}) 
\right]. \label{rate_function}
\end{equation} 
Differentiating (\ref{tmp3}) and (\ref{rate_function}) 
with respect to $\boldsymbol{\mathcal Q}$ and 
$\tilde{\boldsymbol{\mathcal Q}}$, respectively, 
we obtain the following equations giving extrema of (\ref{tmp3}) and 
(\ref{rate_function}): 
\begin{equation} \label{equation1}
\tilde{\boldsymbol{\mathcal Q}}^{\mathrm{s}} 
= -\beta^{-1}(\boldsymbol{I}+\boldsymbol{\Sigma}
\boldsymbol{\mathcal Q}^{\mathrm{s}})^{-1}\boldsymbol{\Sigma},  
\end{equation}
\begin{equation} \label{equation2}
\boldsymbol{\mathcal Q}^{\mathrm{s}} =  
\lim_{K\rightarrow\infty}\frac{1}{K}\sum_{k=1}^{K} 
\frac{1}{{\mathcal M}_{k}^{(u)}
(\tilde{\boldsymbol{\mathcal Q}}^{\mathrm{s}})}
\mathrm{E}_{\boldsymbol{\mathcal X}^{k}}\left[
 \boldsymbol{w}^{k}{\boldsymbol{w}^{k}}^{*}
 \mathrm{e}^{
  \mathrm{tr}\left(
   \tilde{\boldsymbol{\mathcal Q}}^{\mathrm{s}}
   \boldsymbol{w}^{k}{\boldsymbol{w}^{k}}^{*} 
  \right)
 }
\right]. 
\end{equation}
Differentiating (\ref{tmp3}) with respect to $u$ and substituting  
(\ref{equation1}) to it, we can evaluate (\ref{free_energy_replica}) as 
\begin{equation}
{\mathcal F} = 
-\lim_{u\rightarrow 0}\left[
 \beta^{-1}\frac{\partial G^{(u)}}{\partial u}
 (\boldsymbol{\mathcal Q}^{\mathrm{s}}) 
 - \frac{\partial I^{(u)}}{\partial u}
 (\boldsymbol{\mathcal Q}^{\mathrm{s}})  
\right]. 
\end{equation} 

To evaluate the solution of (\ref{equation1}) and (\ref{equation2}) 
in the $u\rightarrow0$ limit 
analytically we assume that the replica symmetry holds, i.e., 
$\boldsymbol{\mathcal Q}^{\mathrm{s}}$ and 
$\tilde{\boldsymbol{\mathcal Q}}^{\mathrm{s}}$ are invariant 
under exchange of non-zero replica indexes. Then,    
$\boldsymbol{\mathcal Q}^{\mathrm{s}}$ and 
$\tilde{\boldsymbol{\mathcal Q}}^{\mathrm{s}}$ can be written as 
\begin{eqnarray}
\boldsymbol{\mathcal Q}^{\mathrm{s}} = 
\begin{pmatrix}
 \boldsymbol{Q}^{0} & \boldsymbol{e}_{u}^{T}\otimes \boldsymbol{M} \\
 \boldsymbol{e}_{u}\otimes\boldsymbol{M}^{*} 
 & \boldsymbol{I}_{u}\otimes (\boldsymbol{Q}^{1}-\boldsymbol{Q}) 
 + \boldsymbol{e}_{u}\boldsymbol{e}_{u}^{T}\otimes\boldsymbol{Q}
\end{pmatrix}, \\
\tilde{\boldsymbol{\mathcal Q}}^{\mathrm{s}} = 
\begin{pmatrix}
 \tilde{\boldsymbol{Q}}^{0} & 
 \boldsymbol{e}_{u}^{T}\otimes\tilde{\boldsymbol{M}} \\
 \boldsymbol{e}_{u}\otimes\tilde{\boldsymbol{M}}^{*} 
 & \boldsymbol{I}_{u}\otimes (\tilde{\boldsymbol{Q}}^{1}
 -\tilde{\boldsymbol{Q}}) 
 + \boldsymbol{e}_{u}\boldsymbol{e}_{u}^{T}\otimes
 \tilde{\boldsymbol{Q}}
\end{pmatrix},
\end{eqnarray}
where $\boldsymbol{M}$, $\tilde{\boldsymbol{M}}$ 
are $N$-by-$N$ matrices and 
$\boldsymbol{Q}^{0}$, $\tilde{\boldsymbol{Q}}^{0}$, 
$\boldsymbol{Q}^{1}$, $\tilde{\boldsymbol{Q}}^{1}$, 
$\boldsymbol{Q}$, and $\tilde{\boldsymbol{Q}}$ are 
$N$-by-$N$ Hermitian matrices. 
By solving (\ref{equation1}) we have in the $u\rightarrow 0$ limit 
\begin{equation}
\tilde{\boldsymbol{Q}}^{0} = 
\boldsymbol{0}, 
\
\tilde{\boldsymbol{M}} = 
\boldsymbol{R}^{-1},
\ 
\tilde{\boldsymbol{Q}}^{1} = 
\tilde{\boldsymbol{Q}} 
- \tilde{\boldsymbol{M}}, 
\
\tilde{\boldsymbol{Q}} = 
\boldsymbol{R}^{-1}\boldsymbol{R}_{0}\boldsymbol{R}^{-1}, \label{solution2}
\end{equation}
where $\boldsymbol{R}_{0}$, $\boldsymbol{R}$ are defined as 
\begin{eqnarray}
\boldsymbol{R}_{0} &=& 
\sigma^{2}\boldsymbol{I}_{N} 
+ \beta(\boldsymbol{Q}^{0}-\boldsymbol{M}-\boldsymbol{M}^{*}
+\boldsymbol{Q}), \label{R_0_tmp} \\ 
\boldsymbol{R} &=& 
\sigma^{2}\boldsymbol{I}_{N} 
+ \beta(\boldsymbol{Q}^{1}-\boldsymbol{Q}). \label{R_tmp}
\end{eqnarray}
It is straightforward to confirm that $\boldsymbol{R}_{0}$, 
$\boldsymbol{R}$ are positive definite. 

We move on to calculating the moment generating 
function~(\ref{moment_generating}). 
$\boldsymbol{R}_{0}$ can be decomposed 
into the product of two nonsingular matrices, i.e., $\boldsymbol{R}_{0}=
\sqrt{\boldsymbol{R}_{0}}\sqrt{\boldsymbol{R}_{0}}^{*}$. 
From (\ref{solution2}) the moment generating 
function~(\ref{moment_generating}) is evaluated as
\begin{equation}
{\mathcal M}_{k}^{(u)}(\tilde{\boldsymbol{\mathcal Q}}^{\mathrm{s}}) =
\mathrm{E}_{\boldsymbol{\mathcal X}^{k}}\left[
 \mathrm{e}^{
  \left\| 
   \boldsymbol{b}^{k} 
  \right\|^{2} 
  - \sum_{\alpha =0}^{u}(\boldsymbol{H}^{k}\boldsymbol{x}^{k,\alpha})^{*}
  \boldsymbol{R}_{\alpha}^{-1}\boldsymbol{H}^{k}\boldsymbol{x}^{k,\alpha} 
 }
\right],
\end{equation} 
where $\boldsymbol{R}_{\alpha}=\boldsymbol{R}$ for $\alpha=1, \ldots, u$, 
and where $\boldsymbol{b}^{k}$ is defined as
\begin{equation} 
\boldsymbol{b}^{k} = 
(\sqrt{\boldsymbol{R}_{0}})^{-1}\boldsymbol{H}^{k}\boldsymbol{x}^{k,0} 
+ \sqrt{\boldsymbol{R}_{0}}^{*}\boldsymbol{R}^{-1}\sum_{\alpha =1}^{u}
\boldsymbol{H}^{k}\boldsymbol{x}^{k,\alpha}.
\end{equation}
By applying the transform: 
\begin{equation}
\mathrm{e}^{\|\boldsymbol{b}^{k}\|^{2}} = 
\int \frac{1}{\pi^{N}\det\boldsymbol{R}_{0}}
\mathrm{e}^{
 -{\boldsymbol{y}^{k}}^{*}\boldsymbol{R}_{0}^{-1}\boldsymbol{y}^{k} 
 + 2\Re({\boldsymbol{b}^{k}}^{*}(\sqrt{\boldsymbol{R}_{0}})^{-1}
 \boldsymbol{y}^{k})
}
\boldsymbol{dy}^{k}, 
\end{equation}
we obtain 
\begin{eqnarray}
{\mathcal M}_{k}^{(u)}(\tilde{\boldsymbol{\mathcal Q}}^{\mathrm{s}}) =
\int \mathrm{E}_{\boldsymbol{x}^{k}}\left[
 p(\boldsymbol{y}^{k}|\boldsymbol{x}^{k}, \boldsymbol{H}^{k}; 
 \boldsymbol{R}_{0})
\right]
\nonumber \\
\left\{
 \frac{
  \mathrm{E}_{\boldsymbol{x}^{k}}\left[
   p(\boldsymbol{y}^{k}|\boldsymbol{x}^{k}, \boldsymbol{H}^{k}; 
   \boldsymbol{R})
  \right]
 }
 {
  p(\boldsymbol{y}^{k}|\boldsymbol{0}, \boldsymbol{H}^{k}; \boldsymbol{R})
 }  
\right\}^{u}\boldsymbol{dy}^{k},   
\end{eqnarray}
where $p(\boldsymbol{y}^{k}|\boldsymbol{x}^{k}, 
\boldsymbol{H}^{k}; \boldsymbol{R})$ is defined as
\begin{equation}
p(\boldsymbol{y}^{k}|\boldsymbol{x}^{k}, \boldsymbol{H}^{k}; 
\boldsymbol{R}) = 
\frac{\mathrm{e}^{ 
 - (\boldsymbol{y}^{k}-\boldsymbol{H}^{k}\boldsymbol{x}^{k})^{*}
 \boldsymbol{R}^{-1}
 (\boldsymbol{y}^{k}-\boldsymbol{H}^{k}\boldsymbol{x}^{k})
}}{\pi^{N}\det\boldsymbol{R}}. 
\end{equation}

Assuming that $\boldsymbol{R}_{0}$ is equal to $\boldsymbol{R}$, 
from (\ref{equation2}) we obtain 
\begin{equation}
\boldsymbol{Q}^{0}-\boldsymbol{M}-\boldsymbol{M}^{*}+\boldsymbol{Q} =
\boldsymbol{Q}^{1}-\boldsymbol{Q} =
\lim_{K\rightarrow\infty}\frac{1}{K}\sum_{k=1}^{K}
\boldsymbol{\mathcal E}^{k}, \label{multiuser_effect} 
\end{equation}
\begin{equation} \label{error_covariance_matrix}
\boldsymbol{\mathcal E}^{k} =
\mathrm{E}\left[
 \boldsymbol{H}^{k}\left(
  \boldsymbol{x}^{k} 
  - \langle \boldsymbol{x}^{k} \rangle_{\mathrm{MIMO}_{1}} 
 \right)
 \left(
  \boldsymbol{x}^{k} 
  - \langle \boldsymbol{x}^{k} \rangle_{\mathrm{MIMO}_{1}}
 \right)^{*}
 {\boldsymbol{H}^{k}}^{*}
\right], 
\end{equation} 
where $\langle \cdot \rangle_{\mathrm{MIMO}_{1}}$ is defined as
\begin{equation} \label{MIMO_posterior_mean_estimator}
\left\langle 
 \boldsymbol{x}^{k}
\right\rangle_{\mathrm{MIMO}_{1}} = 
\frac{
 \mathrm{E}_{\boldsymbol{x}^{k}}\left[ 
  \boldsymbol{x}^{k}
  p(\boldsymbol{y}^{k} |\boldsymbol{x}^{k}, \boldsymbol{H}^{k}; 
  \boldsymbol{R})
 \right]
}
{
 \mathrm{E}_{\boldsymbol{x}^{k}}\left[
  p(\boldsymbol{y}^{k} |\boldsymbol{x}^{k}, \boldsymbol{H}^{k}; 
  \boldsymbol{R})
 \right]
}.
\end{equation}
Note that $\boldsymbol{\mathcal E}^{k}$ is averaged with 
respect to $\boldsymbol{H}^{k}$ due to the law of large numbers.
Substituting (\ref{multiuser_effect}) to (\ref{R_0_tmp}) and 
(\ref{R_tmp}), we obtain the fixed-point equation  
\begin{equation} \label{R}
\boldsymbol{R} = 
\sigma^{2}\boldsymbol{I}_{N} 
+ \beta\lim_{K\rightarrow\infty}\frac{1}{K}\sum_{k=1}^{K}
\boldsymbol{\mathcal E}^{k}, 
\end{equation}
which is the extension of the Tse-Hanly equation 
\cite{Tse99} to the case of the MIMO-CDMA channel. 

Calculating $G^{(u)}(\boldsymbol{\mathcal{Q}}^{\mathrm{s}})$, 
$I^{(u)}(\boldsymbol{\mathcal{Q}}^{\mathrm{s}})$ and differentiating them 
with respect to $u$, we can evaluate $\mathcal{F}$ as  
\begin{eqnarray}
\frac{\beta}{\ln2}{\mathcal F} = 
\lim_{K\rightarrow\infty}\frac{1}{K}\sum_{k=1}^{K}\beta M_{k}
{\mathcal C}_{\mathrm{MIMO}_{1}}^{k}
+ N\log(\pi\sigma^{2}\mathrm{e}) \nonumber \\
+ \mathrm{KL}\left(
 \mathcal{CN}\left(\boldsymbol{0}, \sigma^{2}\boldsymbol{I}_{N}\right)||
 \mathcal{CN}\left(\boldsymbol{0}, \boldsymbol{R}\right)
 \right), \label{expect_free_energy_rs} 
\end{eqnarray}
where $\mathrm{KL}(\cdot ||\cdot )$ represents the Kullback-Leibler 
divergence, and where $\mathcal{C}_{\mathrm{MIMO}_{1}}^{k}$ is defined as 
\begin{equation} \label{spectral_efficiency_MIMO}
{\mathcal C}_{\mathrm{MIMO}_{1}}^{k} = 
\frac{1}{M_{k}}\mathrm{E}\left\{ 
 \left. 
 \log\frac{
 p(\boldsymbol{y}^{k}|\boldsymbol{x}^{k}, \boldsymbol{H}^{k}; 
 \boldsymbol{R})
 }
 {
  \mathrm{E}_{\boldsymbol{x}^{k}}\left[
   p(\boldsymbol{y}^{k}|
    \boldsymbol{x}^{k}, \boldsymbol{H}^{k}; \boldsymbol{R})
  \right]
 }
 \right| \boldsymbol{H}^{k}
\right\}.
\end{equation}
Note that (\ref{expect_free_energy_rs}) depends only on $\boldsymbol{R}$. 
In the case that there exist multiple solutions of (\ref{R}), 
one should choose the solution achieving the supremum of (\ref{tmp3}) 
in a neighborhood of $u=0$, i.e., the solution minimizing 
(\ref{expect_free_energy_rs}). 

From the above mentioned analysis we claim that the spectral 
efficiency of the MIMO-CDMA channel is evaluated as 
\begin{eqnarray}
{\mathcal C}_{\mathrm{MIMO-CDMA}} = 
\lim_{K\rightarrow\infty}\frac{1}{K}\sum_{k=1}^{K}
\frac{\beta M_{k}}{\bar{M}}\mathrm{E}_{\boldsymbol{H}^{k}}\left[
 {\mathcal C}_{\mathrm{MIMO}_{1}}^{k}
\right] \nonumber \\
+ \frac{1}{\bar{\mu}N}\mathrm{KL}\left(
 \mathcal{CN}\left(\boldsymbol{0}, \sigma^{2}\boldsymbol{I}_{N}\right)||
 \mathcal{CN}\left(\boldsymbol{0}, \boldsymbol{R}\right)
 \right). \label{spectral_efficiency_rs}
\end{eqnarray}

$\mathrm{E}_{\boldsymbol{H}^{k}}[{\mathcal C}_{\mathrm{MIMO}_{1}}^{k}]$ 
can be interpreted as the spectral efficiency of the following single-user 
MIMO Gaussian channel for the $k$th user:
\begin{equation} \label{MIMO_channel}
\boldsymbol{y}^{k} = \boldsymbol{H}^{k}\boldsymbol{x}^{k} 
+ \boldsymbol{n}^{k}, 
\quad
\boldsymbol{n}^{k}\sim\mathcal{CN}(\boldsymbol{0},\boldsymbol{R}). 
\end{equation} 
From the replica analysis for the moment sequence of the posterior 
mean estimator  
$\langle x_{m_{k}}^{k} \rangle_{\mathrm{MIMO-CDMA}}=
\mathrm{E}[x_{m_{k}}^{k}|\vec{\boldsymbol{y}},\boldsymbol{
\mathcal{S}},\boldsymbol{\mathcal{H}}]$, 
(\ref{spectral_efficiency_rs}) indicate that the MIMO-CDMA channel
with the MMSE detector front end is decoupled into a bank of 
single-user MIMO channels with the MMSE detector 
front ends in the many-user limit.  
It is easy to extend this decoupling result to the cases of 
the linear MMSE detector or the matched filter. 

\section{Replica Analysis of MIMO}
So far we have not specified statistics of the elements of 
$\boldsymbol{H}^{k}$. In order to obtain a more concrete expression   
of (\ref{spectral_efficiency_rs}),  
we consider a fading model with a limited number of scatterers 
\cite{Muller03}
\begin{equation}
\boldsymbol{H}^{k} = {\boldsymbol{\Phi}^{k}}^{*}\boldsymbol{A}^{k}
\boldsymbol{\Theta}^{k},
\end{equation}
where $\boldsymbol{\Theta}^{k}$ is a $S_{k}$-by-$M_{k}$ steering 
matrix which describes the propagation from the transmit antennas 
of the $k$th user to $S_{k}$ scattering objects between the $k$th 
user and the receiver, where $\boldsymbol{A}^{k}=\mathrm{diag}(A_{1}^{k}, 
\ldots, A_{S_{k}}^{k})$ is a $S_{k}$-by-$S_{k}$ diagonal matrix 
which accounts for attenuation at the scattering objects between 
the $k$th user and the receiver, and where $\boldsymbol{\Phi}^{k}$ is a 
$S_{k}$-by-$N$ steering matrix which describes the propagation from 
the scattering objects between the $k$th user and the receiver to the 
receive antennas of the receiver. 
We assume that the elements of $\boldsymbol{\Phi}^{k}$ and 
$\boldsymbol{\Theta}^{k}$ are i.i.d. random variables with the variances 
$1/N$ and $1/S_{k}$, respectively, and that the elements of 
$\boldsymbol{A}^{k}$ are independent random variables 
subject to the normalization 
$\mathrm{E}[\mathrm{tr}(\boldsymbol{A}^{k}{\boldsymbol{A}^{k}}^{*})]=S_{k}$.

It is difficult to evaluate (\ref{error_covariance_matrix}) 
and (\ref{spectral_efficiency_MIMO}) analytically except in special 
cases, e.g., the data symbol vector of the $k$th user $\boldsymbol{x}^{k}$ 
follows a circularly symmetric complex Gaussian distribution or 
the users and the receiver have a few numbers of the antennas. 
Thus, we evaluate (\ref{error_covariance_matrix}) and 
(\ref{spectral_efficiency_MIMO}) using the replica method in the 
many-antenna limit where $M_{k}$, $S_{k}$, and $N$ tend to infinity 
with their ratios $\rho_{k}=S_{k}/N$, $\gamma_{k}=M_{k}/S_{k}$ fixed. 
One might think that the assumption of the many-antenna limit 
is impractical, but it can be a good approximate approach to systems with 
a few antennas if the elements of $\boldsymbol{x}^{k}$ and 
$\boldsymbol{H}^{k}$ follow circularly symmetric complex Gaussian 
distributions \cite{Tse05}.  
On the other hand, the spectral efficiency is not invariant 
under exchange of the order of the many-user limit and the many-antenna 
limit. In the MIMO-CDMA literature, however, it may be reasonable to take 
the many-user limit first.  

By regarding ${\boldsymbol{\Phi}^{k}}^{*}$ and 
$\boldsymbol{A}^{k}\boldsymbol{\Theta}^{k}\boldsymbol{x}^{k}$ as 
the channel matrix and the data symbol vector, respectively, 
we can evaluate the expectation of (\ref{spectral_efficiency_MIMO}) as 
\begin{eqnarray} \label{spectral_efficiency_MIMO_rs}
\lim\mathrm{E}_{\boldsymbol{H}^{k}}\left[
 \mathcal{C}_{\mathrm{MIMO}_{1}}^{k}
\right]
= \lim_{S_{k},M_{k}\rightarrow\infty}
\mathrm{E}_{\boldsymbol{A}^{k}\boldsymbol{\Theta}^{k}}\left[
 \mathcal{C}_{\mathrm{MIMO}_{2}}^{k} 
\right] \nonumber \\
+ \lim_{N\rightarrow\infty}\frac{1}{\mu_{k}N}
\mathrm{KL}\left(
 \mathcal{CN}(\boldsymbol{0}, \boldsymbol{R})||
 \mathcal{CN}(\boldsymbol{0}, \boldsymbol{W}^{k})
\right), 
\end{eqnarray}
where $\mu_{k}=M_{k}/N$, where $\lim$ represents the many-antenna limit, 
and where $\mathcal{C}_{\mathrm{MIMO}_{2}}^{k}$ is the spectral efficiency 
of the following MIMO channel for the $k$th user: 
\begin{equation} \label{MIMO_channel_2}
\tilde{\boldsymbol{y}}^{k} = \boldsymbol{A}^{k}\boldsymbol{\Theta}^{k}
\boldsymbol{x}^{k} + \tilde{\boldsymbol{n}}^{k}, 
\quad 
\tilde{\boldsymbol{n}}^{k} \sim \mathcal{CN}(\boldsymbol{0},
{\zeta^{k}}^{2}\boldsymbol{I}_{S_{k}}).  
\end{equation}
${\zeta^{k}}^{2}$ and $\boldsymbol{W}^{k}$ satisfy the following 
fixed-point equations: 
\begin{equation} \label{zeta}
{\zeta^{k}}^{-2} = 
\lim_{N\rightarrow\infty}\frac{1}{N}\mathrm{tr}\left[
 {\boldsymbol{W}^{k}}^{-1}
\right],
\; 
\boldsymbol{W}^{k} = 
\boldsymbol{R}
+ \rho_{k}{\mathcal E}_{\mathrm{MIMO}_{2}}^{k}\boldsymbol{I}_{N}, 
\end{equation} 
\begin{equation} \label{mean_squared_error_MIMO}
{\mathcal E}_{\mathrm{MIMO}_{2}}^{k} =
\lim_{M_{k}, S_{k}\rightarrow\infty}
\frac{1}{S_{k}}\mathrm{E}\left[
 \left\|
  \boldsymbol{A}^{k}\boldsymbol{\Theta}^{k}
  (\boldsymbol{x}^{k} 
  - \langle \boldsymbol{x}^{k} \rangle_{\mathrm{MIMO}_{2}})
 \right\|^{2}
\right], 
\end{equation}
where $\langle\boldsymbol{x}^{k}\rangle_{\mathrm{MIMO}_{2}}=
\mathrm{E}[\boldsymbol{x}^{k}|\tilde{\boldsymbol{y}}^{k}, 
\boldsymbol{A}^{k}\boldsymbol{\Theta}^{k}]$ and where the operator 
$\mathrm{E}$ in (\ref{mean_squared_error_MIMO}) represents 
the expectation with respect to 
$p(\tilde{\boldsymbol{y}}^{k},\boldsymbol{x}^{k},
\boldsymbol{A}^{k}\boldsymbol{\Theta}^{k})$. 
In the case that there exist multiple solutions of (\ref{zeta}) 
and (\ref{mean_squared_error_MIMO}),  
one should choose the solution minimizing the spectral 
efficiency~(\ref{spectral_efficiency_MIMO_rs}).

To evaluate (\ref{error_covariance_matrix}) we define a quantity 
$\tilde{\mathcal{F}}_{\mathrm{MIMO}_1}^{k}$ as  
\begin{equation}
\tilde{{\mathcal F}}_{\mathrm{MIMO}_1}^{k} = 
\lim\frac{1}{S_{k}}\ln
\mathrm{E}\left[
 \tilde{Z}_{\mathrm{MIMO}_1}^{(u)}(\boldsymbol{y}^{k}, 
 \boldsymbol{H}^{k}; \boldsymbol{\Omega})
\right], 
\end{equation}
\begin{equation} 
\tilde{Z}_{\mathrm{MIMO}_1}^{(u)} = 
\mathrm{E}_{{\boldsymbol{\mathcal X}}^{k}}\left\{
 \mathrm{e}^{
  \mathrm{tr}\left[
   \boldsymbol{\Omega}^{T}\boldsymbol{F}
  \right]
 }
 \prod_{\alpha=0}^{u}p(\boldsymbol{y}^{k}|\boldsymbol{x}^{k,\alpha}, 
 \boldsymbol{H}^{k};\boldsymbol{R})
\right\}, 
\end{equation}
where $\boldsymbol{F}$ is defined as 
\begin{equation}
\boldsymbol{F}({\boldsymbol{\mathcal X}}^{k},\boldsymbol{H}^{k}) 
= 
S_{k}\boldsymbol{H}^{k}(\boldsymbol{x}^{k} - \boldsymbol{x}^{k,1})
(\boldsymbol{x}^{k} - \boldsymbol{x}^{k,2})^{*}
{\boldsymbol{H}^{k}}^{*}.   
\end{equation}
Then, we obtain in the many-antenna limit \cite{Tanaka02}
\begin{equation} \label{nishimori_formula}
\lim\boldsymbol{\mathcal{E}}^{k} = 
\lim_{u\rightarrow 0}\left.
 \frac{\partial}{\partial\Omega}\tilde{\mathcal{F}}_{\mathrm{MIMO}_1}
\right|_{\boldsymbol{\Omega}=\boldsymbol{0}}.  
\end{equation}
Calculating the right-hand side of (\ref{nishimori_formula}), 
we can evaluate (\ref{error_covariance_matrix}) as 
\begin{equation}
\lim\boldsymbol{\mathcal{E}}^{k} = 
\boldsymbol{R}
- \boldsymbol{R}\tilde{\boldsymbol{W}}^{k^{-1}}\boldsymbol{R}, 
\end{equation}
where $\tilde{\boldsymbol{W}}^{k}$ is a solution of 
(\ref{zeta}) and (\ref{mean_squared_error_MIMO}). 
We have added a tilde to $\boldsymbol{W}^{k}$ in order to make clear that 
$\tilde{\boldsymbol{W}}^{k}$ need not be equal to 
$\boldsymbol{W}^{k}$ in (\ref{spectral_efficiency_MIMO_rs}).
In the case that there exist multiple $\tilde{\boldsymbol{W}}^{k}$  
one should choose the solution minimizing 
(\ref{spectral_efficiency_MIMO_rs}).

Furthermore, we can evaluate 
$\mathrm{E}[\mathcal{C}_{\mathrm{MIMO}_2}^{k}]$ and 
$\mathcal{E}_{\mathrm{MIMO}_2}^{k}$ by applying the above mentioned 
method again. On the assumption that 
$\{x_{m_{k}}^{k}; m_{k}=1,\ldots,M_{k}\}$ are mutually independent, 
$\mathrm{E}[\mathcal{C}_{\mathrm{MIMO}_2}^{k}]$ in the many-antenna limit 
is given by 
\cite{Muller03,Wen06} 
\begin{eqnarray} \label{spectral_efficiency_MIMO2_rs}
\lim_{S_{k},M_{k}\rightarrow\infty}
\mathrm{E}[\mathcal{C}_{\mathrm{MIMO}_2}^{k}]
= 
\lim_{M_{k}\rightarrow\infty}\frac{1}{M_{k}}
\sum_{m_{k}=1}^{M_{k}}\mathcal{C}_{\mathrm{AWGN}}^{k,m_{k}} 
\nonumber \\
+ \lim_{S_{k}\rightarrow\infty}\frac{\gamma_{k}^{-1}}{S_{k}}
\mathrm{E}_{\boldsymbol{A}^{k}}\left[
 \mathrm{KL}\left(
  \mathcal{CN}(\boldsymbol{0}, {\zeta^{k}}^{2}\boldsymbol{I}_{S_{k}})||
  \mathcal{CN}(\boldsymbol{0}, \boldsymbol{\Xi}^{k})
 \right)
\right], 
\end{eqnarray}
where $\mathcal{C}_{\mathrm{AWGN}}^{k,m_{k}}$ is the spectral 
efficiency of the scalar Gaussian channel
\begin{equation} \label{AWGN_channel}
y_{m_{k}}^{k} = x_{m_{k}}^{k} + n_{m_{k}}^{k}, 
\quad
n_{m_{k}}^{k}\sim \mathcal{CN}(0,{\xi^{k}}^{2}). 
\end{equation}
${\xi^{k}}^{2}$ and $\boldsymbol{\Xi}^{k}$ satisfy the following 
fixed-point equations: 
\begin{equation} \label{xi}
{\xi^{k}}^{-2} = 
\lim_{S_{k}\rightarrow\infty}\frac{1}{S_{k}}\sum_{s_{k}=1}^{S_{k}}
\mathrm{E}\left[
 \frac{|A_{s_{k}}^{k}|^{2}}
 {
  {\zeta^{k}}^{2} + \gamma_{k}{\mathcal E}_{\mathrm{AWGN}}^{k}
  |A_{s_{k}}^{k}|^{2}
 }
\right], 
\end{equation}
\begin{equation} \label{Xi}
\boldsymbol{\Xi}^{k} = 
{\zeta^{k}}^{2}\boldsymbol{I}_{S_{k}}
+ \gamma_{k}{\mathcal E}_{\mathrm{AWGN}}^{k}
\boldsymbol{A}^{k}{\boldsymbol{A}^{k}}^{*}, 
\end{equation} 
\begin{equation} \label{mean_squared_error_AWGN}
{\mathcal E}_{\mathrm{AWGN}}^{k} =
\lim_{M_{k}\rightarrow\infty}
\frac{1}{M_{k}}\sum_{m_{k}=1}^{M_{k}}\mathrm{E}\left[
 \left|
  (x_{m_{k}}^{k} 
  - \langle x_{m_{k}}^{k} \rangle_{\mathrm{AWGN}})
 \right|^{2}
\right], 
\end{equation}
where $\langle x_{m_{k}}^{k}\rangle_{\mathrm{AWGN}}
=\mathrm{E}[x_{m_{k}}^{k}|y_{m_{k}}^{k}]$. 
In the case that there exist multiple solutions of (\ref{xi}), 
(\ref{Xi}), and (\ref{mean_squared_error_AWGN}),  
one should choose the solution minimizing the spectral 
efficiency~(\ref{spectral_efficiency_MIMO2_rs}).

The result of a replica analysis claims that the moment sequence of 
$\langle x_{m_{k}}^{k}\rangle_{\mathrm{MIMO-CDMA}}$ converges 
to the moment sequence of 
$\langle x_{m_{k}}^{k} \rangle_{\mathrm{AWGN}}$ in the many-user 
and many-antenna limits. 
Since the MMSE detector is information lossless in the scalar 
Gaussian channel, $\mathcal{C}_{\mathrm{AWGN}}^{k,m_{k}}$ can be  
interpreted as the spectral efficiency of the MMSE detector 
in the MIMO-CDMA channel.  

On the other hand, (\ref{mean_squared_error_MIMO}) is 
evaluated as 
\begin{equation} \label{E_MIMO2_rs}
\mathcal{E}_{\mathrm{MIMO}_{2}}^{k} 
= 
\lim_{S_{k}\rightarrow\infty}\frac{1}{S_{k}}\sum_{s_{k}=1}^{S_{k}}
\mathrm{E}\left[
 {\zeta^{k}}^{2} - \frac{{\zeta^{k}}^{4}}{{\zeta^{k}}^{2} + \gamma_{k}
 \tilde{\mathcal{E}}_{\mathrm{AWGN}}^{k}|A_{s_{k}}^{k}|^{2}}
\right], 
\end{equation}
where $\tilde{\mathcal{E}}_{\mathrm{AWGN}}^{k}$ is a solution of 
(\ref{xi}), (\ref{Xi}), and (\ref{mean_squared_error_AWGN}). 
In the case that there exist multiple 
$\tilde{\mathcal{E}}_{\mathrm{AWGN}}^{k}$  
one should choose the solution minimizing 
(\ref{spectral_efficiency_MIMO2_rs}). 
We can obtain the same fixed-point equations for 
${\zeta^{k}}^{2}$ and ${\xi^{k}}^{2}$ by means of calculating the 
expectation of $\mathcal{C}_{\mathrm{MIMO}_{1}}^{k}$ with respect to 
$\boldsymbol{\Theta}^{k}$ and then evaluating the asymptotic 
distribution of singular values of $\boldsymbol{\Phi}^{k}$. 

We can easily confirm that the spectral 
efficiency~(\ref{spectral_efficiency_rs}) coincides with the 
spectral efficiency of the i.i.d. Rayleigh fading MIMO-CDMA channel 
in the case that scattering is very rich. 
In the $\gamma_{k}\rightarrow 0$ limit with $\mu_{k}=\rho_{k}\gamma_{k}$ 
fixed, (\ref{xi}) reduces to ${\xi^{k}}^{2}={\zeta^{k}}^{2}$.  
Expanding (\ref{E_MIMO2_rs}) with respect to $\gamma_{k}$ and 
substituting it to (\ref{zeta}), we obtain  
\begin{equation} \label{zeta_Rayleigh}
{\zeta^{k}}^{-2} = 
\lim_{N\rightarrow\infty}\frac{1}{N}\mathrm{tr}\left[
 {\boldsymbol{W}^{k}}^{-1}
\right],
\; 
\boldsymbol{W}^{k} = \boldsymbol{R} + \mu_{k}
\mathcal{E}_{\mathrm{AWGN}}({\zeta^{k}}^{2})\boldsymbol{I}_{N}. 
\end{equation} 
(\ref{zeta_Rayleigh}) coincides with the fixed-point equation for 
${\zeta^{k}}^{2}$ and $\boldsymbol{W}^{k}$ in the case of the i.i.d. 
Rayleigh fading. Hence, (\ref{spectral_efficiency_rs}) converges 
to the spectral efficiency of the i.i.d. Rayleigh fading MIMO-CDMA 
channel in the $\gamma_{k}\rightarrow 0$ limit.  

The above mentioned results indicate that the MIMO 
channel~(\ref{MIMO_channel}) with the MMSE detector front end 
is decoupled into the bank of scalar Gaussian channels with the 
MMSE detector front ends in the many-antenna 
limit even if the number of the scatterers is limited. 
It is easy to extend this decoupling result to the cases of 
the linear MMSE detector or the matched filter. 

\section{Numerical evaluation}
\begin{figure}[htbp]
 \includegraphics[scale=0.46]{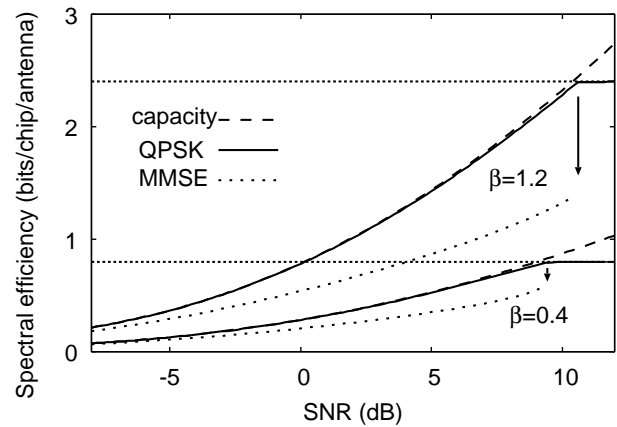}
 \caption{
  Spectral efficiency of the MIMO-CDMA channel versus SNR. 
  $\mu=1.0$ and $\gamma=1.0$.
 }
 \label{snr}
\end{figure}

\begin{figure}[htbp]
 \includegraphics[scale=0.46]{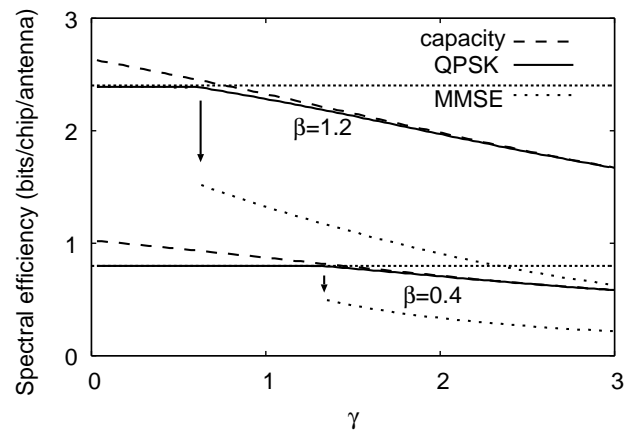}
 \caption{
  Spectral efficiency of the MIMO-CDMA channel for QPSK modulation 
  versus $\gamma$. $\mu=1.0$ and SNR $=10$ dB. 
  The spectral efficiency coincides with the spectral efficiency of 
  the i.i.d. Rayleigh fading MIMO-CDMA channel in the 
  $\gamma \rightarrow 0$ limit.  
 }
 \label{gamma}
\end{figure}

To evaluate the spectral efficiency numerically we consider the case 
that $\rho=\rho_{k}$, $\gamma=\gamma_{k}$, 
$\{x_{m_{k}}^{k}\}$ are i.i.d. random variables with the 
variance $P=\mathrm{E}[|x_{m_{k}}^{k}|^{2}]$, and  
$|A_{s_{k}}^{k}|^{2}=1$ with probability $1$.   
$\mu=\rho\gamma$ represents the ratio of the number of transmit 
antennas to the number of receive antennas, i.e., $\mu=M/N$.
We denote the received signal-to-noise ratio per transmit antenna 
by SNR $=P/\sigma_{0}^{2}$. 
Figure~\ref{snr} displays the spectral 
efficiency~(\ref{spectral_efficiency_rs}) versus SNR for QPSK 
modulation (4-QAM). We find that the spectral efficiency of the 
MIMO-CDMA channel with QPSK modulation is very close to the capacity 
but there exists a large gap between the spectral efficiency of the 
MMSE detector and the capacity in the highly loaded system. 
The degradation of the spectral efficiency due to 
the decrease of scatterers is shown in Fig.~\ref{gamma}. 
In this case, an interesting observation is that the spectral efficiency 
does not degrade so much compared with the rich scattered environment 
($\gamma\rightarrow 0$) even when the number of scatterers 
is comparable with the number of the antennas.

\section{Conclusion}
We evaluated the spectral efficiency of the uplink of the MIMO-CDMA 
channel using the replica method. The main result is that the hierarchical 
decoupling principle holds in the MIMO-CDMA channel, i.e., the MIMO-CDMA 
channel is decoupled into the bank of single-user MIMO channels in the 
many-user limit. The resulting single-user MIMO channel is further 
decoupled into the bank of scalar Gaussian channels in the many-antenna 
limit for the fading model with a limited number of scatterers.   
We found numerically that the spectral efficiency of the MIMO-CDMA 
channel with QPSK modulation is very close to the capacity but 
there exists a large gap between the spectral efficiency of the MMSE 
detector and the capacity in a highly loaded system.  

\section*{Acknowledgment}
The authors are grateful for supports from Grant-in-Aid for Scientific 
Research on Priority Areas 18079010, MEXT, Japan. 

\bibliographystyle{junsrt}
\bibliography{isit2007}


\end{document}